\documentstyle[twocolumn,epsf]{jpsj}
\def\runtitle{
NMR Relaxation Rate for 1D Multicomponent
Spin-Orbital Systems
}
\def\runauthor{
Akira {\sc Kawaguchi}, Tatsuya {\sc Fujii} 
and Norio {\sc Kawakami}
} 
\title{NMR Relaxation Rate for One-Dimensional Multicomponent
Spin-Orbital Systems}

\author{Akira {\sc Kawaguchi}, Tatsuya {\sc Fujii} 
and Norio {\sc Kawakami}} 
\inst{Department of Applied Physics, 
Osaka University, Suita, Osaka 565-0871} 

\recdate{}

\abst{
The NMR relaxation rate $1/T_{1}$ is studied for
one-dimensional multicomponent spin-orbital systems.
By combining the bosonization techniques and the 
exact solution of the SU($n$) model, we 
evaluate the power-law exponent and the enhancement factor 
 for  $1/T_1$ at low temperatures.
We discuss how the band splitting affects
the relaxation rate, and find that $1/T_{1}$ may be 
enhanced around the critical value of  the band
splitting. The crossover behavior in $1/T_1$ around the critical point 
is discussed in terms of the low-frequency dynamical spin 
susceptibility.  The effect of the hole-doping 
is also addressed.
}

\kword
{NMR, one-dimension, spin-orbital model, bosonization}

\begin{document}
\sloppy
\maketitle

\section{Introduction}

  Quantum spin systems with orbital degeneracy have attracted considerable 
attention recently, since it was recognized that the orbital degrees 
of freedom  give rise to a variety of interesting phenomena in 
correlated electron systems.
  These hot topics have stimulated theoretical studies on the orbitally 
degenerate Hubbard model and the related multicomponent 
spin-orbital models. 
   As a first step to investigate such orbital effects, the one-dimensional 
(1D) version of  the SU(4) spin-orbital model has been studied numerically
\cite{ueda,troy} and analytically.
\cite{suth,Schlottmann,affleck,itakura,fujii,zhang} 
Such analyses have been also extended to the SU(2)$\times$SU(2) 
spin-orbital model.\cite{pati,azaria,yama,itoi,azaria2,tsukamoto}
However, these studies have been mainly concerned with the phase diagram 
and/or the static properties, which naturally motivates us to 
study dynamical quantities for the above models.

   The NMR measurement may be a powerful probe to study low-energy 
spin dynamics in 1D strongly correlated systems, as demonstrated 
theoretically\cite{schulz,chitra,sachdev1,sachdev2,fujimoto}
and experimentally.\cite{takigawa,chaboussant,Goto}
In this paper, we study the NMR relaxation rate 
for 1D multicomponent spin-orbital systems. For this purpose, we first 
give a  multicomponent generalization of the formula for the 
NMR relaxation rate $1/T_1$ in terms of the bosonization
method.  We then evaluate  the power-law exponent
 and the enhancement factor for $1/T_1$
by exploiting the integrable SU($n$) spin-orbital model.
It is discussed how the band splitting  affects
the relaxation rate. We further study the dynamical spin 
susceptibility to discuss the crossover behavior in $1/T_1$
around the critical point.
The effect of the hole-doping on the relaxation rate
 is also discussed.

This paper is organized as follows. 
In \S 2, we obtain a generalized expression for the NMR relaxation rate,
 and then discuss the 
the effect of the band splitting on $1/T_{1}$ in \S 3.
We present the results for the dynamical spin 
susceptibility in \S 4 to discuss how the crossover behavior
emerges  when the band splitting is changed.
Brief summary is given in \S 5.


\section{Relaxation Rate for Spin-Orbital Systems} 

\subsection{Bosonization approach}

We first derive a generalized expression  for the 
NMR relaxation rate $1/T_{\rm 1}$ for 1D multicomponent electron 
systems. To illustrate the procedure clearly,
let us consider the degenerate Hubbard model with the band splitting,
which possesses $N$-fold orbitals and spins $\sigma=\uparrow,\downarrow$,
\begin{eqnarray}
H=&-&t\sum_{i=1}^{L}\sum_{\sigma}\sum_{n^{'}=1}^{N}
                    (c_{i \sigma n^{'}}^{\dagger}c_{i+1 \sigma n^{'}}+h.c.)
\nonumber \\
&+&\frac{U}{2}\sum_{i=1}^{L}
              \sum_{\sigma \neq \sigma^{'}}
              \sum_{n^{'} \neq n^{''}}
               n_{i \sigma n^{'}}n_{i \sigma^{'} n^{''}}
\nonumber \\
&-&\Delta\sum_{i}\sum_{\sigma}(n_{i}^{(l)}-n_{i}^{(u)}),
\label{eq:hub}
\end{eqnarray}
where $c_{i \sigma n^{'}}^{\dagger}$ is the creation operator for 
an electron with the orbital labeled by $n^{'}(=1,2,\cdots,N)$. 
The band  splitting $\Delta$, which is assumed to be caused
either by the crystal-field effect or by the Zeeman effect,
gives rise to the energetically separated bands with the 
dispersion,  $\varepsilon (k)=-2t\cos ka_0 \pm \Delta$
 where $a_0$ is the lattice spacing. 
We deal with the case that the lower (upper) band is 
 $p$-fold ($q$-fold) degenerate, 
for which the factor 2 due to the spin sector is included.
The  number of electrons is denoted as 
 $n_{i}^{(l)}$ ($n_{i}^{(u)}$) for the lower (upper) band.
 Note that the system possesses SU($2N$) spin-orbital symmetry 
in the absence of the band splitting.\cite{ueda,zhang}
For convenience, we use the number 
$n=2N$ and the index $m=1,2,\cdots,n(=2N)$, 
which specifies the spin as well as the orbital
degrees of freedom on equal footing. 

Passing to  the continuum limit, we now introduce the 
 right- and left-going fermion fields,\cite{voit} 
\begin{equation}
  c_{i m}
   \rightarrow
    \sqrt{a_{0}}\left[
               {\rm e}^{{\rm i}k_{F}x}\psi_{+ m}(x)
               +{\rm e}^{-{\rm i}k_{F}x}\psi_{- m}(x)
                \right],
\label{eq:continuum}
\end{equation}
where the Fermi momentum, $k_{F}$ represents either 
$k_{F}^{(l)}$ for the lower band $(m=1,\cdots p)$ or 
$k_{F}^{(u)}$ for the upper band $(m=p+1,\cdots,n)$. 
These fermion fields are bosonized in terms of 
$\Phi_{m}$ and its dual field $\Theta_{m}$,
\begin{equation}
  \psi_{r m}=\frac{\eta_{r m}}{\sqrt{2\pi a_{0}}}
             \exp(-{\rm i}[r\Phi_{m}-\Theta_{m}]),
\end{equation}
where the boson fields satisfy the commutation relation, 
$[\Phi_{m}(x),\Theta_{m^{'}}(x^{'})]
={\rm i} (\pi/2) \delta_{mm^{'}} {\rm sgn}(x^{'}-x)$. 
The anticommutation relation between fermions with different
species\cite{voit} is preserved by Majorana-fermion operator, $\eta_{r m}$.

Let us now bosonize the Hubbard interaction.
We consider a metallic case, 
for which the charge, spin and orbital excitations are all massless.
In this case, we can neglect both of the backward scattering 
and the Umklapp scattering around the low-energy fixed point.
This simplification reduces the system 
to an effective  multicomponent Tomonaga-Luttinger (TL) model
in external fields,
\begin{eqnarray}
H &=& \sum_{m} \int \frac{{\rm d}x}{2\pi}
            \left\{
                   v_{m}
                    \left[ 
                     \partial_{x} \Phi_{m}(x)
                    \right]^{2}+
                   v_{m}
                    \left[ 
                     \partial_{x} \Theta_{m}(x)
                    \right]^{2}
            \right\} 
\nonumber \\
&&  + \frac{a_0 U}{\pi^{2}}
            \sum_{m \neq m'}\int {\rm d}x
                  \left[ 
                     \partial_{x} \Phi_{m}(x)
                    \right]
                    \left[ 
                     \partial_{x} \Phi_{m'}(x)
                    \right],
\label{eq:boson-eq}
\end{eqnarray}
where $v_m=v_l$ for $m=1,\cdots,p$
and $v_m=v_u$ for $m=p+1,\cdots,n$ are the Fermi velocities 
of elementary excitations.

Note that the Fermi velocities 
in the lower and upper bands are different,
making the diagonalization of the Hamiltonian  (\ref{eq:boson-eq}) 
a little bit complicated. Nevertheless, we can 
perform the diagonalization via the canonical transformation
following the way done for a two-channel TL model.\cite{Nagaosa,Kimura}
To this end, we  introduce the new boson fields, 
$\Phi_{c}$, $\Phi_{\Delta}$ and $\Phi_{\sigma_{m}}$
for the charge, orbital and spin sectors, respectively.
For the lower band, the transformation is given by 
\begin{eqnarray}
 \Phi_{m}&&=\left[
                  \frac{1}{\sqrt{n}}\cos\alpha+
                  \frac{\sqrt{pq}}{p\sqrt{n}}y\sin\alpha
            \right] \Phi_{c}
\nonumber \\ 
 &&        +\left[
                 -\frac{1}{\sqrt{n}}\frac{\sin\alpha}{y}+
                  \frac{\sqrt{pq}}{p\sqrt{n}}\cos\alpha
            \right] \Phi_{\Delta} 
\nonumber \\ 
 &&        +\sum_{m{'}=3}^{p+1}
                  \frac{1}{\sqrt{(m{'}-1)(m'-2)}}
\nonumber \\
 &&            \times \left[
                      \theta(m'-m-1)-(m'-2)\delta_{m',m+1}
                 \right] \Phi_{\sigma_{m'-2}}, 
\label{trans1}
\end{eqnarray}
for $m=1,2,\cdots,p$, whereas for the upper band, 
\begin{eqnarray}
  \Phi_{m}&&=\left[
                   \frac{1}{\sqrt{n}}\cos\alpha-
                   \frac{\sqrt{pq}}{q\sqrt{n}}y\sin\alpha
             \right] \Phi_{c}
\nonumber \\
  &&         +\left[
                   -\frac{1}{\sqrt{n}}\frac{\sin\alpha}{y}-
                    \frac{\sqrt{pq}}{q\sqrt{n}}\cos\alpha
              \right] \Phi_{\Delta} 
\nonumber \\ 
  &&         +\sum_{m'=p+2}^{n}
                    \frac{1}{\sqrt{(m'-p)(m'-p-1)}}
\nonumber \\
  &&         \times \left[
                    \theta(m'-m)-(m'-p-1)\delta_{m',m}
             \right]\Phi_{\sigma_{m'-2}},
\label{trans2}
\end{eqnarray}
for $m=p+1,\cdots,n$.   We thus end up with    
the effective Hamiltonian in the diagonalized form, 
\begin{eqnarray}
&&  H_{c}\! =\! \int \frac{{\rm d}x}{2\pi}
            \left\{
                   \frac{\tilde{v}_{c}}{\tilde{K}_{c}}
                    \left[ 
                     \partial_{x} \Phi_{c}(x)
                    \right]^{2}+
                   \tilde{v}_{c}\tilde{K}_{c}
                    \left[ 
                     \partial_{x} \Theta_{c}(x)
                    \right]^{2}
            \right\}, 
\nonumber \\
&&  H_{\Delta}\! =\! \int \frac{{\rm d}x}{2\pi}
                 \left\{ 
                  \frac{\tilde{v}_{\Delta}}
                  {\tilde{K}_{\Delta}}
                    \left[ 
                     \partial_{x} \Phi_{\Delta}(x)
                    \right]^{2}
                 +\tilde{v}_{\Delta}\tilde{K}_{\Delta}
                   [ \partial_{x} \Theta_{\Delta}(x) ]^{2}
                 \right\}, 
\nonumber \\
&&  H_{\sigma_{\nu}}\! =\!  \int \frac{{\rm d}x}{2\pi} 
                        \left\{
                         \frac{\tilde{v}_{\sigma_{\nu}}}{K_{\sigma_{\nu}}}
                          \left[
                           \partial_{x}\Phi_{\sigma_{\nu}}(x)
                          \right]^{2}
                        +\tilde{v}_{\sigma_{\nu}}K_{\sigma_{\nu}}
                          [
                           \partial_{x}\Theta_{\sigma_{\nu}}(x)
                          ]^{2}
                        \right\}, 
\nonumber \\                        
\label{eq:effective}
\end{eqnarray}
where $\tilde{v}_{\sigma_{\nu}}=\tilde{v}_{l}$ 
for $\nu=1,\ldots,p-2$ and 
$\tilde{v}_{\sigma_{\nu}}=\tilde{v}_{u}$ 
for $\nu=p-1,\ldots,n-2$. 
The explicit formulae for the parameters 
introduced in the transformation as well as the TL parameters 
are listed in the Appendix.
Note that $H_c$ ($H_\Delta$) describes the Gaussian theory 
for charge  excitations (inter-band excitations), for which 
the TL parameter $\tilde{K}_{c}$ ($\tilde{K}_{\Delta}$)
characterizes the U(1) critical line.  On the other hand, spin excitations
in each band are described by $H_{\sigma_{\nu}}$, which still possess
SU($p$) (SU($q$)) symmetry for the  upper (lower) band. 
This fixes the value $K_{\sigma_\nu}=1$.

\subsection{Expression for $1/T_1$ at low temperatures}

We now derive the formula for the NMR relaxation rate, $1/T_{1}$. 
Let us  first write down
the time-dependent spin-spin correlation function 
at  finite temperatures,  $<S^{+}(x,t)S^{-}(0,0)>$,  
\begin{eqnarray}
  &&<S^{+}(x,t)S^{-}(0,0)>
\nonumber \\
   &\simeq&
    \sum_{\kappa} \prod_{\mu}
      \frac{{\rm e}^{{\rm i} \kappa x}
             \left( 
                    \frac{\beta^{'}}{\tilde{v}_{\mu} }
             \right)^{2x_{\mu}} }
           { \left[
                   \sinh \frac{\beta^{'}}{\tilde{v}_{\mu} }
                    (x+\tilde{v}_{\mu}t)
             \right]^{2\Delta_{\mu}^{+}}
             \left[ 
                   \sinh\frac{\beta^{'}}{\tilde{v}_{\mu} }
                    (x-\tilde{v}_{\mu}t)
             \right]^{ 2\Delta_{\mu}^{-} } },
\nonumber \\
\label{correlation}
\end{eqnarray}
with  $\beta^{'}=\pi k_{B}T$, where $\mu (=c,\Delta,\sigma_{\nu})$ 
classifies  massless modes in eq. (\ref{eq:effective}). 
The phase factor ${\rm e}^{{\rm i} \kappa x}$ 
comes from the large momentum transfer across the Fermi points,
 where the momentum $\kappa (=2k_{F}^{(l,u)},k_{F}^{(l)}+ k_{F}^{(u)})$ 
specifies such excitations.
Here $\Delta^{\pm}_{\mu}$ are conformal dimensions which are related
to the scaling dimension $x_{\mu}$ as 
 $ x_{\mu}=\Delta_{\mu}^{+} + \Delta_{\mu}^{-}$.
The dynamical spin susceptibility is expressed in terms of the 
above correlation functions, 
\begin{equation}
  \chi_{\perp}(k,\omega)=
     -{\rm i}\int {\rm d}t {\rm d}x \theta(t)
                  \langle [S^{+}(x,t),S^{-}(0,0)]
                  \rangle
                   {\rm e}^{{\rm i}(\omega t-kx)}.
\label{suscep}
\end{equation}
The transverse spin correlation function, $<S^{z}S^{z}>$, can be expressed
in a similar manner. 
By computing  eqs. (\ref{correlation}) and (\ref{suscep})
with the use of eqs. (\ref{trans1}), (\ref{trans2}) and (\ref{eq:effective}),
we obtain a generic formula for $1/T_{1}$ at low temperatures, 
\begin{eqnarray}
     \frac{1}{T_{1}}&=&\lim_{\omega\rightarrow 0}
                         \frac{2 k_{B}T}{\hbar^{2}\omega}
                          \int\frac{{\rm d}k}{2\pi}
                           A_{\perp}^{2}(k){\rm Im}\chi(k,\omega)
\nonumber \\
                    &\sim&
                       \sum_{\kappa}B_{\kappa}
                        \Gamma_{\kappa}(\Delta)
                         T^{\eta_{\kappa}-1},
\label{nmr}
\end{eqnarray}
where
\begin{eqnarray}
     \Gamma_{\kappa}(\Delta)=
       \left[
             \prod_\mu 
              \left( \frac{1}{\tilde{v}_{\mu}} \right)^{2x_{\mu}}
      \right].
\label{nmr2}
\end{eqnarray}
Here $A_{\perp}$ is hyperfine form factor
and $B_{\kappa}$ is a constant independent of 
the temperature.  The corresponding critical exponents
$\eta_{\kappa}=2 \sum_{\mu} x_{\mu}$ are obtained as, 
\begin{eqnarray}
\eta_{2k_{F}^{(l)}}&& = 2-\frac{2}{p}
                      + \frac{2}{n}\left\{\cos\alpha+
                          \sqrt{\frac{q}{p}}y\sin\alpha\right\}^{2}
                            \tilde{K}_{c}
\nonumber \\ 
    && 
    +\frac{2}{n}\left\{-\frac{\sin\alpha}{y}+
              \sqrt{\frac{q}{p}}\cos\alpha\right\}^{2}
                                   \tilde{K}_{\Delta}, 
       \label{eta2kfl}\\
\eta_{2k_{F}^{(u)}}&& = 2-\frac{2}{q}
                      + \frac{2}{n}\left\{\cos\alpha-
                          \sqrt{\frac{p}{q}}y\sin\alpha\right\}^{2}
                            \tilde{K}_{c}
\nonumber \\ 
    && +\frac{2}{n}\left\{-\frac{\sin\alpha}{y}-
              \sqrt{\frac{p}{q}}\cos\alpha\right\}^{2}
                                   \tilde{K}_{\Delta}, 
       \label{eta2kfu}\\
\eta_{k_{F}^{(l)}+k_{F}^{(u)}}&& = 
     2-\frac{1}{p}-\frac{1}{q}
\nonumber \\
      && + \frac{1}{2n}\left[2\cos\alpha
               +\left(\sqrt{\frac{q}{p}}-\sqrt{\frac{p}{q}}\right)
                   y\sin\alpha\right]^{2}\tilde{K}_{c}
\nonumber \\ 
      && +  \frac{1}{2n}\left(\sqrt{\frac{q}{p}}+\sqrt{\frac{p}{q}}\right)^{2}
                  \left(\frac{\sin\alpha}{y}\right)^{2}\tilde{K}^{-1}_{c}
\nonumber \\
     &&+ \frac{1}{2n}\left[-2\frac{\sin\alpha}{y}
         +\left(\sqrt{\frac{q}{p}}-\sqrt{\frac{p}{q}}\right)
               \cos\alpha\right]^{2}\tilde{K}_{\Delta}
\nonumber \\ 
     &&+ \frac{1}{2n}\left(\sqrt{\frac{q}{p}}+\sqrt{\frac{p}{q}}\right)^{2}
           (\cos\alpha)^{2}\tilde{K}^{-1}_{\Delta}.
 \label{etainter}
\end{eqnarray}
This completes the derivation of the formula for $1/T_1$,
 which generalizes the results of 
Sachdev\cite{sachdev1} to multicomponent 
cases.  It is to be noted here that we have explicitly obtained the 
enhancement factor  $\Gamma_{\kappa}$ in (\ref{nmr}),
which has been neglected in ordinary literatures.
We shall see that this quantity  plays an important role for the 
 NMR relaxation rate  in 1D systems, particularly when the system
is located around the quantum phase transition point
around which the velocities are dramatically renormalized.
 Some explicit examples are shown in the following section.

We have so far treated the model in the continuum limit
within a weak coupling approach, so that it is not
straightforward to evaluate $1/T_1$ as a function of the microscopic
model parameters.  In the following section 
we evaluate  the renormalized 
parameters $\tilde{v}_{\mu}$ and ${\tilde K}_{\mu}$, 
by exploiting the solvable model to deduce characteristic 
behaviors in $1/T_1$.

%
\section{Exact Critical Properties of $1/T_1$}
In this section, we  exactly evaluate the renormalized parameters 
$\tilde{v}_{\mu}$ and ${\tilde K}_{\mu}$ to discuss
critical properties of $1/T_1$ for the 
SU($n$) spin-orbital chain with band splitting. Also, the 
effect of hole-doping is discussed.

\subsection{Spin-orbital systems}

Let us start with the spin-orbital system in a
insulating phase. In the limit $U \rightarrow \infty$, 
the spin-orbital sector in the degenerate Hubbard model 
(\ref{eq:hub}) is reduced to the 
SU($n$) antiferromagnetic Heisenberg model (each site
is occupied by one electron), which is written
down  in terms of the fermion operators,
\begin{eqnarray}
H_J \! = \! J \! \sum_{i}^{L}
        \! \sum_{\sigma,\sigma^{'}}
        \! \sum_{n^{'},n^{''}}^{N}
        \!   (c_{i \sigma n^{'}}^{\dag} c_{i \sigma' n^{''}} 
            c_{i+1 \sigma^{'} n^{''}}^{\dag} c_{i+1 \sigma n^{'}}
              -n_{i}n_{i+1}),
\nonumber \\
\label{sunmodel}
\end{eqnarray}
where only the  single occupation  of electrons is allowed 
at each site. The band splitting $\Delta$ due to, e.g. a 
crystalline field, forms 
two energetically separated bands, which are composed of 
the $p$ lower  and $q$ upper  bands including 
spin degrees of freedom, $n=2N=p+q$. 

The exact solution of the model (\ref{sunmodel})
was obtained by Sutherland,\cite{suth} and its properties 
have been clarified thus far.\cite{ueda,troy,itakura,fujii,zhang}
Following standard procedures, we can evaluate the 
velocities of elementary excitations and  the critical 
exponents by applying finite-size scaling techniques\cite{belavin,cardy} 
to the excitation spectrum.\cite{suth}
 
In the insulating phase,  charge excitations are
frozen,  while  spin excitations are still gapless which
possess SU($p$) (SU($q$)) symmetry for the lower (upper) band.
 By applying the finite-size scaling\cite{cardy} to the excitation 
spectrum, we obtain the TL parameter as,
\begin{eqnarray}
 \tilde{K}_{\Delta}=\frac{n}{pq}\xi_{\Delta(p,q)}^{2}, \hspace{2mm} 
 \alpha =0,
\label{eq:relation1}
\end{eqnarray}
which generalizes the results of the SU(4) case.\cite{itakura}
In the above expression, 
the dressed charge\cite{izergin,woynarovich,frahm,kawakami}
$\xi_{\Delta(p,q)}\equiv
\xi_{\Delta(p,q)}(\lambda_{p}^{0})$,
which features the critical line of the TL liquid, 
is given by the integral equation, 
\begin{eqnarray}
\xi_{\Delta(p,q)}(\lambda_p) \! = \! 1
     \! +\! \int_{-\lambda_{p}^{0}}^{\lambda_{p}^{0}}
             G(\lambda_{p}-\lambda'_{p})
              \xi_{\Delta(p,q)}(\lambda'_{p}){\rm d}\lambda'_{p},
\end{eqnarray}
where $G(\lambda)$ is defined as 
\begin{eqnarray}
&& G(\lambda) = F_{p,p}(\lambda)+F_{q,q}(\lambda)
     -\frac{1}{2\pi}\int_{-\infty}^{\infty}x^{2}
    {\rm e}^{{\rm i}k\lambda}{\rm d}k,
\\
&& F_{\mu,\nu}(\lambda) = \frac{1}{2\pi}\int_{-\infty}^{\infty}
        x\frac{x^{\mu-1}-x^{1-\mu}}{x^{\nu}-x^{-\nu}}
            {\rm e}^{{\rm i}k\lambda}{\rm d}k,
\end{eqnarray}
with $x=\exp(-|k|/2)$. 
Here the cut-off parameters $\pm\lambda_{p}^{0}$ are determined
by minimizing the free energy for a given band splitting  $\Delta$.
\cite{suth}

Having determined the TL parameter $\tilde{K}_{\Delta}$, 
it is now straightforward to derive the critical exponents for 
various correlation functions by exploiting the scaling relations 
obtained  in eqs.(\ref{eta2kfl})-(\ref{etainter}). Note that the 
TL parameter $\tilde{K}_{c}$ for the charge degree of freedom 
vanishes in the  insulating system. 
Let us discuss the longitudinal  spin susceptibility
$\chi_{\perp}$, as an example.
For excitations with  momentum transfer $2k^{(l,u)}_F$ 
within  the lower and upper bands, 
 we  obtain  the exponents $\eta_{2k_F^{(l,u)}}$
from (\ref{eta2kfl}) and (\ref{eta2kfu}), 
\begin{eqnarray}
 \eta_{2k_F^{(l)}}&=&
                  2\left[1-\frac{1}{p}\right]
                    +\frac{2q}{np}\tilde{K}_{\Delta},
\label{eq:2kfl} \\
 \eta_{2k_F^{(u)}}&=&
                  2\left[1-\frac{1}{q}\right]
                    +\frac{2p}{nq}\tilde{K}_{\Delta}.
\label{eq:2kfu}
\end{eqnarray}
On the other hand, for the momentum transfer 
$k^{(l)}_F \pm k^{(l)}_F$ we  use the 
relation (\ref{etainter}), which yields
\begin{eqnarray}
 \eta_{k^{(l)}_F + k^{(l)}_F}
      =&& \left[2-\frac{1}{p}-\frac{1}{q}\right]
              +\frac{n}{2pq}\tilde{K}^{-1}_{\Delta}
 \cr
       &+& \frac{1}{2n}\left(
                    \sqrt{\frac{q}{p}}-\sqrt{\frac{p}{q}}
                           \right)^2
            \tilde{K}_{\Delta}.
 \label{eq:2kfl-l}
\end{eqnarray}
The scaling relations (\ref{eq:2kfl}), (\ref{eq:2kfu})
and (\ref{eq:2kfl-l}), which are the generalizations of 
the ordinary TL relations to multicomponent cases, 
characterize critical properties of the 
degenerate model with band splitting.

We note here that for the relaxation via excitations around the 
zero momentum transfer, the critical exponent is fixed to
the canonical value, resulting in  the specific formula for this
relaxation channel,
\begin{eqnarray}
  \frac{1}{T_{1}} 
         \sim \left( \frac{1}{v} \right)^{2} T,
\label{Korringa}
\end{eqnarray}
which holds for any value of the band splitting, although the velocity
$v$ may be  changed according to the splitting. Note that
the quantity $1/v^2$, which  originates from the enhancement factor  
$\Gamma_{\kappa}$ in (\ref{nmr}), is proportional to the square of the 
density of states for elementary excitations. Therefore,
this expression is regarded as a 1D analogue of the  Korringa 
relation.\cite{korringa}

\begin{figure}[htb]
\begin{center}
\vspace{-0cm}
\leavevmode \epsfxsize=85mm 
\epsffile{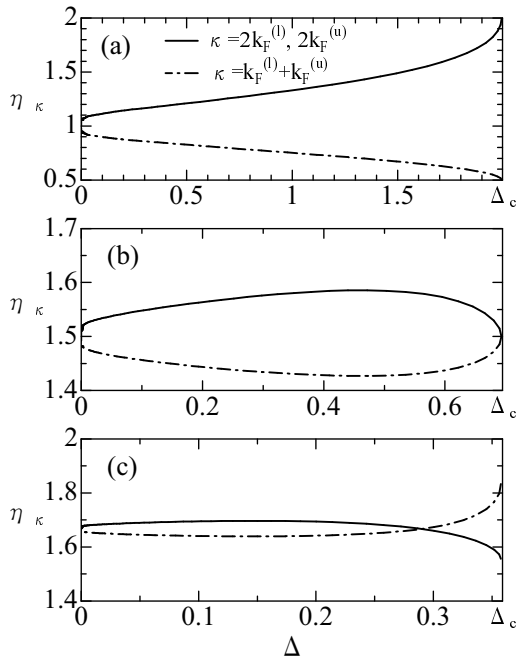}
\vspace{-3.0cm}
\end{center}
\caption{Critical exponents of the NMR relaxation rate  
in the case of SU($n$) spin-orbital chain  in which 
$n$-fold multiplet split into two set of  $n/2$-fold  states
with the band splitting $\Delta$: (a)$n=2$ (b)4, (c)6.
Note that for the $n=2,6$ cases, $\Delta$ may be regarded as the 
magnetic field. The critical value $\Delta_c$ at which
the upper band becomes massive  is given by
(a) $1.93$, (b) $0.69$ and (c) $0.36$, respectively. 
}
\label{fig:2}
\end{figure}

\begin{figure}[htb]
\begin{center}
\vspace{-0cm}
\leavevmode \epsfxsize=80mm 
\epsffile{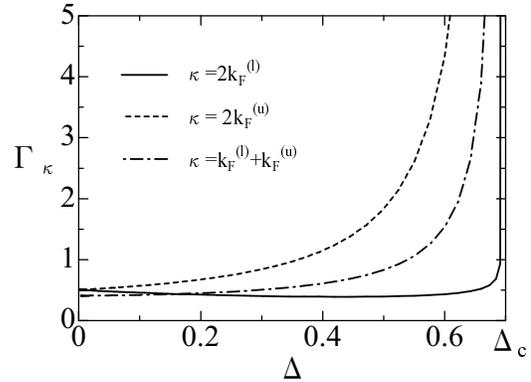}
\vspace{-6.0cm}
\end{center}
\caption{The enhancement factor $\Gamma_{\kappa}$ in $1/T_{1}$
for the SU(4) chain as a function of the band splitting $\Delta$.
The critical field at which the upper band becomes massive 
is $\Delta_{c}=\log 2$.
}
\label{fig:1}
\end{figure}

Let us now focus on the specific case
for which SU($n$) multiplet splits into two SU($n/2$) bands ($n=2N$). 
Note that the $n=4$ case has been intensively studied thus far. 
\cite{ueda,troy,itakura,fujii,zhang}
 In Fig. \ref{fig:2} we show the critical exponents 
as a function of the band splitting  $\Delta$.
In the absence of the band splitting, $\Delta=0$, 
the TL parameter $\tilde{K}_{\Delta}$
is given by $\tilde{K}_{\Delta}=1$, which results in 
$\eta_{2k_{F}}=2(1-1/n)$.\cite{affleck}
Hence the power-law temperature dependence  becomes less singular
as the orbital degeneracy is increased. Although
 the  critical exponents vary continuously
when $\Delta$ is increased up to its critical value
$\Delta_c$, their values do not change so 
dramatically  to induce the qualitative difference
in the temperature dependence except for $n=2$ (single
orbital case). On the other hand,
beyond $\Delta_c$, the upper band 
becomes massive, being separated from the lower band. 
Thus for $\Delta>\Delta_c$, 
the low-energy physics can be described by the SU($n/2$)
spin model. Correspondingly, the critical exponent 
shows a discontinuity at $\Delta_c$.  For example, 
$\eta_{2k_{F}^{(l)}}$ abruptly changes from $2(1-1/p+1/p^2)$ to 
$2(1-1/p)$  for the SU($n$) model at $\Delta=\Delta_c$.
As claimed by Yamashita and Ueda,\cite{ueda} this discontinuity
does not imply that the relaxation rate itself possesses such singularity,
but there instead occurs the crossover behavior in $1/T_1$.
This point is discussed in more detail in the next section.

We  now observe how the enhancement factor  $\Gamma _{\kappa}$ 
in  (\ref{nmr}) behaves as a function of the band splitting.
The computed results for $\Gamma_{\kappa}$ 
 are  shown in Fig. \ref{fig:1}. 
As $\Delta$ is increased, $\Gamma_{\kappa}$
is increased monotonically and is divergently enhanced near 
$\Delta=\Delta_c$.  This characteristic behavior is caused  by 
the renormalization of the velocities, which gives rise to
the edge singularity in density of states. 
This may be regarded as a 1D analog of the "mass enhancement" well known
 for the Fermi liquid in three dimension.  Note that  $\Gamma_{0}$ as well as 
$\Gamma_{2k_{F}^{(l)}}$ within the lower band still has the finite values
even for $\Delta >\Delta_c$, whereas the others 
should vanish after divergently increased at $\Delta_c$.
As mentioned before, such a singularity is apparent, and 
should be replaced by the crossover behavior.
We note that the above enhancement factor has been 
usually neglected  to analyze  the temperature dependence 
of $1/T_1$,  However, when the 
{\it field-dependence} of $1/T_1$ is studied at low temperatures, 
this should be properly taken into account, in particular
around the critical point  $\Delta=\Delta_c$.


\subsection{Hole-doped systems}

We now turn to a hole-doped case. 
For a metallic system with finite holes, we consider  
the multicomponent {\it t-J} model,
\begin{eqnarray}
H =-t\sum_{i=1}^{L}\sum_{\sigma}\sum_{n^{'}=1}^{N}
                 (c_{i \sigma n^{'}}^{\dagger}c_{i+1 \sigma n^{'}}+h.c.)
+ H_J.
\end{eqnarray}
For the supersymmetric case, $\it t=J$, the exact solution was obtained,
\cite{suth,Schlottmann} which will be used here. Though this model is 
rather special in its appearance, some essential properties 
for a doped system can be described well, as is shown below.

In Fig. \ref{eta-hole}, we show the critical exponents 
for the SU($n$) symmetric case 
without the  band-splitting $\Delta=0$ ($n=2,4,6$). 
In this case, the critical exponent 
$\eta(=\eta_{2k_{F}^{(l)}}=\eta_{2k_{F}^{(u)}}
=\eta_{k_{F}^{(l)}+k_{F}^{(u)}})$  is
\begin{eqnarray}
 \eta= 2-\frac{2}{n}+\frac{2}{n}\tilde{K}_{c}, \;\;\;
  \tilde{K}_{c}=\frac{1}{n}\xi_{h(n)}^{2},
\label{tjexp}
\end{eqnarray}
where the dressed charge $\xi_{h(n)}\equiv
\xi_{h(n)}(\lambda_{n}^{0})$ is given by 
\begin{eqnarray}
\hspace{-0.5cm}
\xi_{h(n)}(\lambda_n)\! =\! 1
     \! +\! \int_{-\lambda_{n}^{0}}^{\lambda_{n}^{0}}
         F_{n,n}(\lambda_{n}-\lambda'_{n})
         \xi_{\delta(n,n)}(\lambda'_{n}){\rm d}\lambda'_{n}.
\end{eqnarray}
We note that the relation (\ref{tjexp})  generalizes
 the TL scaling relation to  SU($n$) electron systems.
The computed results are shown in Fig.\ref{eta-hole}. 
When the hole concentration is increased ($\mu$ is decreased), 
the critical exponent for the NMR relaxation rate,
$\eta$, is monotonically increased from  $2(1-1/n+1/n^2)$ up to 
$2$ (band bottom), because $\tilde{K}_{c}=1/n \rightarrow 1$
in this case.
We are now discussing a rather special solvable model, so that
the global feature of the critical exponent should depend  
on the model. However, characteristic properties close to
the insulating phase ($\mu \simeq \mu_c$) is expected 
to be rather general 
 because in the small hole-doping region the charge sector is 
scaled to the strong coupling fixed point (hard-core bosons) 
irrespective of the bare value of $J$.

\begin{figure}[htb]
\begin{center}
\vspace{-0cm}
\leavevmode \epsfxsize=80mm 
\epsffile{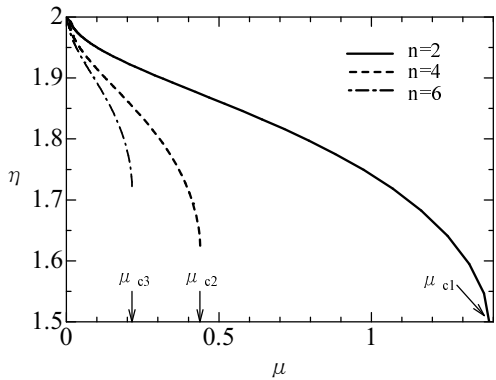}
\vspace{-6.0cm}
\end{center}
\caption{Critical exponents of the NMR relaxation rate 
for the SU($n$) supersymmetric {\it t-J} model ($n=2,4,6$)
as a function of the chemical potential $\mu$.
At $\mu=\mu_c$, the system
becomes the Mott insulator:
$\mu_{c1}\sim 1.39$, $\mu_{c2}\sim 0.44$ 
and $\mu_{c3}\sim 0.22$, respectively for $n=2,4,6$.
}
\label{eta-hole}
\end{figure}

\begin{figure}[htb]
\begin{center}
\vspace{-0cm}
\leavevmode \epsfxsize=85mm 
\epsffile{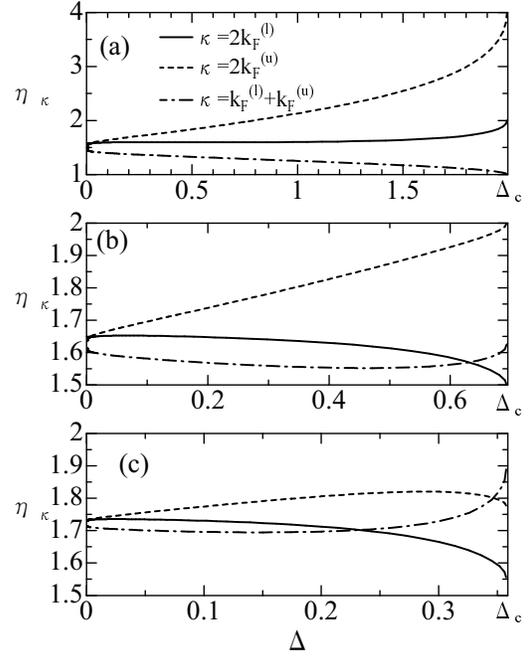}
\vspace{-3.0cm}
\end{center}
\caption{Critical exponents as a function of 
the band splitting $\Delta$ 
in a metallic system close to the Mott insulating phase: 
(a) SU(2), (b) SU(4) and  (c) SU(6).
}
\label{fig:3}
\end{figure}
To see such  properties for the 
critical exponents, we now focus on the hole-doped
case close to insulator ($\mu \sim \mu_c$). 
In this limiting case,  the TL parameters in (\ref{eta2kfl}), (\ref{eta2kfu}) 
and (\ref{etainter}) are reduced to,
\begin{eqnarray}
\tilde{K}_{\Delta}=\frac{n}{pq}\xi_{\Delta(p,q)}^{2},
\;\tilde{K}_{c}=\frac{1}{n},\;
\cos\alpha=1,
\nonumber \\
y\sin\alpha=-\sqrt{\frac{p}{q}}\left(
1-\frac{n}{p}\xi_{B(p,q)}\right),\;
\frac{1}{y}\sin\alpha=0,
\end{eqnarray}
where in addition to the dressed charge (\ref{eq:relation1}),
 we have introduced another dressed charge 
$\xi_{B(p,q)}\equiv\xi_{B(p,q)}(\lambda_{n}^{0})$
which incorporates the interference between the charge and spin 
degrees of freedom,
\begin{eqnarray}
\xi_{B(p,q)}(\lambda_n)=
  \int_{-\lambda_{p}^{0}}^{\lambda_{p}^{0}}
F_{2,q}(\lambda_{n}-\lambda'_{p})
  \xi_{B(p,q)}(\lambda'_{p}){\rm d}\lambda'_{p}.
\end{eqnarray}

In Fig. \ref{fig:3}, the critical exponents in
a metallic system with the small hole 
concentration are shown as  a
function of the band splitting $\Delta$ ($n=2,4,6$). 
As soon as holes are doped into the insulator, all the exponents 
are increased (compare Fig.\ref{fig:2} with Fig.\ref{fig:3}),
because the charge sector couples with the spin sector, and thus
affects  the relaxation process via "gauge interaction"
due to Fermi statistics.
For the limiting case, $\Delta\rightarrow \Delta_c$, 
the critical exponents approach
\begin{eqnarray}
 \eta_{2{k_F}^{(l)}}&=&
                   2-\frac{4}{n}+\frac{8}{n^2},
\\
 \eta_{2{k_F}^{(u)}}&=&
                   2-\frac{4}{n}+\frac{16}{n^2},
\\ 
 \eta_{{k_F}^{(l)}+{k_F}^{(u)}}&=&
                   2-\frac{4}{n}+\frac{1}{2}+\frac{2}{n^2}. 
\end{eqnarray}
Beyond this critical band-splitting, the NMR relaxation rate 
is given by that for the SU($n/2$) {\it t-J} model
without band-splitting.  Thus the critical exponent is 
discontinuously changed to smaller values 
at $\Delta=\Delta_c$ for any case of $n$.

We wish to mention that the coefficient $\Gamma_{\kappa}$  
is enhanced for small  hole-doping,
since the  charge velocity $\tilde{v}_{c}$ in eq.(\ref{nmr2}) 
is very small in this region. However, it may not be easy to
observe this enhancement due the renormalized charge velocity, because 
the hole-doping itself increases the value of the exponents, thus 
having a tendency to hide singular behaviors at low temperatures.


\section{Dynamical Spin Susceptibility}

We have shown that the critical exponents as well
as the enhancement factor exhibit singularities around 
the critical point $\Delta_c$. As already mentioned, such singularities 
may not be observed in NMR measurements, but should be replaced by 
the crossover behavior.\cite{ueda}
In order to clearly see how the crossover behavior emerges,
we discuss the dynamical spin susceptibility ${\rm Im}\chi(k,\omega)$
in the low frequency regime.

By exploiting  the velocities and the TL parameters determined by the
exact solution, we have computed  ${\rm Im}\chi(k,\omega)$ at
finite temperatures by using the expression (\ref{suscep}).  
To be specific, we restrict ourselves to the SU(4) spin-orbital
model with the band splitting.

\begin{figure}[htb]
\begin{center}
\leavevmode \epsfxsize=80mm 
\epsffile{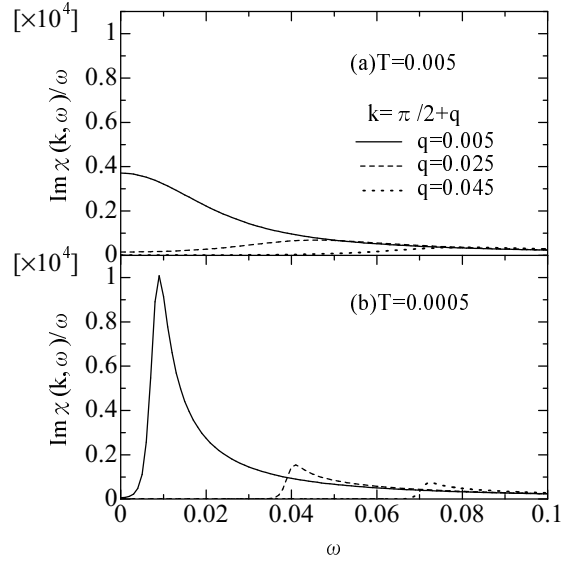}
\vspace{-3.0cm}
\end{center}
\caption{Spectral function  
${\rm Im}\chi(k,\omega)/\omega$
of the SU(4) model ($\Delta=0$) as a function of
the frequency for several values of the momenta $k(=\pi/2+q$):
(a) higher temperature case ($T=0.005$) and 
(b) lower temperature case ($T=0.0005$).
Note that the momentum  $q$ measures the difference from 
$\pi/2(=2k_F^{(l)}=2k_F^{(u)}=k_F^{(l)}+k_F^{(u)})$ around which
elementary excitations become massless.
}
\label{su4-l-h}
\end{figure}

In  Fig. \ref{su4-l-h},
 the spectral function ${\rm Im}\chi(k,\omega)/\omega$ is shown 
for $\Delta=0$ at finite temperatures. 
For all the momenta shown in Fig. \ref{su4-l-h} (a), 
the broad maximum structure is found in the spectral function, 
indicating the presence of overdamped excitations. 
This is because we are dealing with the case of
higher temperatures in (a), $\hbar v q \le k_B T$ ($v=\pi/2$).
On the other hand, if the temperature is decreased with $q$ being fixed, 
the sharp peak structure is developed  around $\omega=\hbar v q$ 
in the spectral function (see (b) for which $\hbar v q\gg k_B T$)
which indicates the presence of propagating spinons with the 
above dispersion relation. 
These behaviors for the propagating mode are analogous
to those observed for the SU(2) spin chain.\cite{sachdev1}
Note that although there are three kinds of massless modes,
all the velocities are the same for the SU(4) symmetric 
case with $\Delta=0$, so that we can observe only one spinon 
mode in the spectral function ${\rm Im}\chi(k,\omega)/\omega$.
The integration over $q$ in the case of (b) determines the
low-temperature power-law behavior of 
$1/T_1$ discussed in the previous section.

\begin{figure}[tb]
\begin{center}
\vspace{-0cm}
\hspace{-1.0cm}
\leavevmode \epsfxsize=95mm 
\epsffile{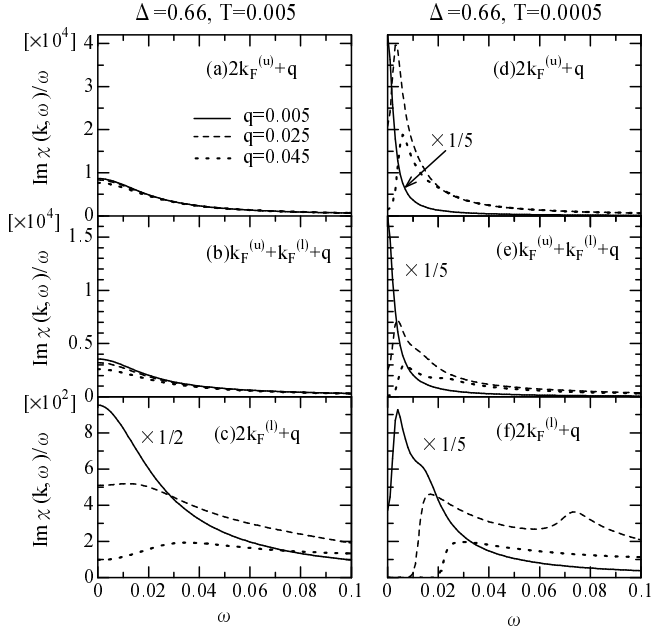}
\vspace{-3.0cm}
\end{center}
\vspace{-1.0cm}
\caption{Spectral function ${\rm Im}\chi(k,\omega)/\omega$
for the SU(4) model with the band splitting 
$\Delta=0.66$ (close to the critical value $\Delta_c\simeq 0.69$).
There are three massless modes, whose velocities are
respectively given by 
 $\tilde{v}_l\simeq2.87$, $\tilde{v}_{\Delta}\simeq0.48$ 
and $\tilde{v}_u\simeq0.10$. 
Note that the values of some solid lines have been scaled down 
by the factor of 1/5 or 1/2. 
}
\label{low-high}
\end{figure}

We next discuss the case with the finite band splitting close to 
its critical value $\Delta_c$, where we have encountered the 
singularities in the NMR  relaxation rate. 
In Fig. \ref{low-high}, we display 
the spectral function ${\rm Im}\chi(k,\omega)/\omega$ 
by choosing two typical values of the temperature.
There are three dominant low-energy excitations with
different velocities around the momenta
$2k_F^{(u)}$, $k_F^{(l)}+k_F^{(u)}(=\pi/2)$  and $2k_F^{(l)}$. 
At higher temperatures  shown in panels (a), (b) and (c), 
we can see the broadened maximum structure at low
frequencies, which is caused by overdamped spin
excitations, as mentioned above.
When the temperature is lowered, each excitation starts
to show different behaviors in the spectral function.
We begin with the spectral function for the $2k_F^{(u)}$ mode 
in (d), and the $k_F^{(l)}+k_F^{(u)}$ mode in (e).
By comparing them with those for the $\Delta=0$ case,
we find that the peak structure in the spectral weight is shifted 
to the extremely low frequency regime, and its peak-position 
is monotonically increased with the increase of the momentum $q$,
reflecting the  fact that the velocities of these  excitations
become very small around $\Delta_c$.

Recall here that the NMR relaxation rate (\ref{nmr}) is 
determined by the integration of the spectral function over 
the momentum in the limit $\omega\rightarrow 0$. 
In the previous section, we have shown  that
$1/T_1$ is extremely enhanced by the velocity renormalization
around $\Delta_c$.  However, we should keep in mind the following
things. We are now considering the continuum limit of the 
original lattice model, so that when the system is 
close to the critical point $\Delta_c$, the energy range where such a
 treatment may be valid becomes pretty small.  For instance,
in the above case, the energy range where the dispersion relation 
$\omega\simeq\hbar \tilde{v}_u q$ holds 
should be in  $-k_F^{(u)}\sim q\sim k_F^{(u)}$ 
(the cutoff is given by $k_F^{(u)}\simeq 0.15$ for $\Delta=0.66$). 
Therefore the enhancement of $1/T_1$ 
can be seen in  the very low frequency regime 
within the small cutoff parameter.  Hence,
even if such enhancement is naively expected to be observed at low
temperatures, it may be obscured when the temperature is increased
beyond the above cutoff energy.

In contrast to the above two modes, the spectral function for the 
 $2k_F^{(l)}$ mode in Fig. \ref{low-high}(f) shows quite different 
behaviors; it develops two peaks of 
propagating spinons around $2k_F^{(l)}$, whose dispersions are given by 
$\omega\simeq\hbar \tilde{v}_{\Delta} q$ and 
$\omega\simeq\hbar \tilde{v}_l q$, respectively. 
When the system is close to the critical point, the interband 
excitations with the dispersion $\tilde{v}_{\Delta} q$
becomes less important. On the other hand, in such cases,
the excitation of $\omega\simeq\hbar \tilde{v}_l q$
within the lower band becomes more dominant
to control the relaxation process. In fact, beyond $\Delta_c$,
this excitation completely determines the $1/T_1$ at low temperatures.

From the above analyses  of the  dynamical spin susceptibility, 
we can see how the enhancement of $1/T_1$
in the vicinity of the critical field naturally crossovers
to the region where $1/T_1$ is dominated by excitations in
 the lower band, when the energy or the temperature is raised.


\section{Summary}

We have studied  the NMR relaxation rate
$1/T_{1}$ for 1D  multicomponent spin-orbital  systems.
 By generalizing the bosonization approach 
of Sachdev, we have obtained a generalized formula of $1/T_{1}$ 
for multicomponent quantum systems with the band splitting.
We have then exactly estimated the relaxation rate 
by using the SU($n$) integrable model for the insulating 
as well as metallic phases.  It has been found that 
the power-law  temperature dependence of $1/T_1$  becomes 
less singular as the orbital degeneracy is increased.
This is also the case for the finite band splitting, as far
as $\Delta$ is smaller than its critical value $\Delta_c$.
We have also pointed out that the relaxation rate may be enhanced 
around $\Delta_c$ due to the dramatic renormalization of the velocities. 
 This type of the enhancement may be common to 1D quantum systems
when the spin/orbital gap is formed or collapsed by external fields.
In particular, this effect becomes important to analyze the
{\it field-dependence} of $1/T_1$ at low temperatures.
It may be quite interesting to experimentally 
observe the enhancement in  $1/T_1$, or the corresponding 
crossover behavior at very low temperatures in 
the NMR experiments.

\section*{Acknowledgements}
We would like to thank Y. Tsukamoto for valuable discussions.
This work was partly supported by a Grant-in-Aid from the Ministry
of Education, Science, Sports and Culture of Japan.
A. K. and F. T. were supported by 
Japan Society for the Promotion of Science. 
\section*{Appendix}

The  parameters for the canonical 
transformation in (\ref{trans1}) and (\ref{trans2}) are obtained  as, 
\begin{eqnarray}
&&  \tan 2\alpha = \frac{\delta v}{v_{0}(y-\gamma y^{-1})}
                    \frac{4\sqrt{pq}}{n}, \hspace{2mm}
         y^{2} = \frac{K_{c}^{-2}+\gamma}{\gamma K_{\Delta}^{-2}}, 
\nonumber \\
&&         v_{0} = \frac{1}{n}(pv_{l}+qv_{u}), 
\hspace{2mm}         
      \delta v = \frac{1}{2}(v_{l}-v_{u}), 
\nonumber \\
&&        \gamma = \frac{qv_{l}+pv_{u}}{pv_{l}+qv_{u}},
\nonumber
\end{eqnarray}
where $v_l$ and $v_u$ are the bare velocities, and the TL parameters
$K_c$ and $K_{\Delta}$ are given by 
\begin{eqnarray}
  K_c &=& \left(1+\frac{2(n-1)a_0 U}{\pi v_0}\right)^{-1/2}, 
  \nonumber \\
  K_{\Delta} &=& \left(1-\frac{2(n-1)a_0 U}{\pi v_0}\right)^{-1/2}.
  \nonumber
\end{eqnarray}
 The above set of the parameters in the transformation 
gives the diagonalized  Hamiltonian in the text,
for which the renormalized velocities, 
$\tilde{v}_{c(\Delta )}$ and the
TL parameters, $\tilde{K}_{c(\Delta)}$ read, 
\begin{eqnarray}
  &&\tilde{v}^{2}_{c {\Delta}} 
    = \frac{4pq}{N^{2}} \delta v^{2}
      +\frac{1}{2}v_{0}^{2}
        [ 
         (K_{c}^{-2}+\gamma^{2}K_{\Delta}^{-2})
  \pm  T(\alpha)], 
\nonumber \\
  &&\tilde{K}^{2}_{c(\Delta)}
    =y^{\mp 2}
  \frac{
          [K_{c}^{-2}+\gamma^{2}
            K_{\Delta}^{-2}+2\gamma]
           \pm
          T(\alpha)
         }
         { 
          [K_{c}^{-2}(1+2\gamma K_{\Delta}^{-2})
          +\gamma^{2}K_{\Delta}^{-2}]
          \pm
           T(\alpha)
         },
\nonumber
\end{eqnarray}
where $T(\alpha)=(K_{c}^{-2}-\gamma^{2}K_{\Delta}^{-2})\sqrt{1+\tan 2\alpha}$.
%


%


\begin{thebibliography}{99}
%

\bibitem{ueda}
Y. Yamashita, N. Shibata, and K. Ueda: 
Phys. Rev. B {\bf 58} (1998) 9114. 
%
%
\bibitem{troy}
B. Frischmuth, F. Mila, M. Troyer: 
Phys. Rev. Lett. {\bf 82} (1999) 835. 

\bibitem{suth}
B. Sutherland: 
Phys. Rev. B {\bf 12} (1975) 3795.
%
\bibitem{Schlottmann}
P.Schlottmann: 
Phys. Pev. Lett. {\bf 69} (1992) 2396. 
%
\bibitem{affleck}
I. Affleck: 
Nucl. Phys. B {\bf 265} (1986) 409. 
%
\bibitem{itakura}
T. Itakura and N. Kawakami:
J. Phys. Soc. Jpn. {\bf 64} (1995) 2321. 
%
\bibitem{fujii}
T. Fujii, Y. Tsukamoto and N. Kawakami: 
J. Phys. Soc. Jpn. {\bf 68} (1999) 151.
%
\bibitem{zhang} 
Y. Q. Li, M. Ma, D. N. Shi and F. C. Zhang: 
Phys. Rev. B {\bf 60} (1999) 12781.

\bibitem{pati}
S. K. Pati and R. R. P. Singh: 
Phys. Rev. Lett. {\bf 81} (1998) 5406.
%
\bibitem{azaria}
P. Azaria, A. O. Gogolin, P. Lecheminant and A. A. Nersesyan: 
Phys. Rev. Lett. {\bf 83} (1999) 624.
%
\bibitem{yama}
Y. Yamashita, N. Shibata, and K. Ueda: 
cond-mat/9908237.
%
\bibitem{itoi}
C. Itoi, S. Qin and I. Affleck: 
cond-mat/9910109.
%
\bibitem{azaria2}
P. Azaria, E. Boulat and P. Lecheminant: 
cond-mat/9910218.
%
\bibitem{tsukamoto}
Y. Tsukamoto, N. Kawakami, Y. Yamashita and K. Ueda: 
Physica B in press. 
%
%


%
%
%
%


\bibitem{schulz}
H. J. Schulz: 
Phys. Rev. B {\bf 34} (1986) 6372.
%
\bibitem{chitra}
R. Chitra and T. Giamarchi: 
Phys. Rev. B {\bf 55} (1997) 5816.
%
\bibitem{sachdev1}
S. Sachdev: 
Phys. Rev. B {\bf 50} (1994) 13006.
%
\bibitem{sachdev2}
S. Sachdev: 
Phys. Rev. Lett. {\bf 78} (1997) 943; 
Phys. Rev. B {\bf 57} (1998) 8307.
%
\bibitem{fujimoto}
S. Fujimoto: 
cond-mat/9905415. 
%
\bibitem{takigawa}
M. Takigawa, N. Motoyama, H. Eisaki and S. Uchida: 
Phys. Rev. Lett. {\bf 76} (1996) 4612; 
Phys. Rev. B {\bf 56} (1997) 13681.
%
%
%
\bibitem{chaboussant}
G. Chaboussant, Y. Fagot-Revurat, M.-H. Julien, M. E. Hanson, 
C. Berthier, M. Horvatic, L. P. L{\'e}vy and O. Piovesana: 
Phys. Rev. Lett. {\bf 80} (1998) 2713. 
%
\bibitem{Goto}
T. Goto, Y. Fujii, Y. Shimaoka, T. Maekawa and J. Arai: 
Physica B in press. 
%

\bibitem{voit}
See for example, J. Voit: 
Rep. Prog. Phys. {\bf 58} (1995) 977. 
%
%
\bibitem{Nagaosa}
N. Nagaosa and T. Ogawa: 
Solid State Commun. {\bf 88} (1993) 295.
%
\bibitem{Kimura}
T. Kimura, K. Kuroki, and H. Aoki: 
Phys. Rev. B {\bf 53} (1996) 9572.
%


\bibitem{belavin}
A. A. Belavin, A. M. Polyakov and A. B. Zamolodchikov: 
Nucl. Phys. B {\bf 241} (1984) 333.
%
\bibitem{cardy}
J. L. Cardy: 
Nucl. Phys. B {\bf 270} (1986) 186.
%
%

%
\bibitem{izergin}
A, G. Izergin, V. E. Korepin and N. Yu. Reshetikhin: 
J. Phys. A {\bf 22} (1989) 2615. 
%
\bibitem{woynarovich}
F. Woynarovich: 
J. Phys. A {\bf 22} (1989) 4243.
%
\bibitem{frahm}
H. Frahm and, V. E. Korepin: 
Phys. Rev. B {\bf 42} (1990) 10553.
%
\bibitem{kawakami}
N. Kawakami and S.-K. Yang:  
Phys. Rev. Lett. {\bf 65} (1990) 2309.
%
\bibitem{korringa}
J. Korringa: 
Physca {\bf 16} (1950) 601.

\end{thebibliography}
\end{document}